\documentstyle [12pt,epsf]{article}

\input{epsf}
\begin{document}
\begin{titlepage}
\begin{center}

{\Large \bf Casimir Energy of Massless Fermions in the Slab-bag}\\
\vspace{.3in}
{\large\em  R.D.M. De $Paola$\footnote[1]{e-mail: rpaola@lafex.cbpf.br},
            R.B. $Rodrigues$\footnote[2]{e-mail: robson@lafex.cbpf.br},
        and N.F. $Svaiter$\footnote[3]{e-mail: nfuxsvai@lafex.cbpf.br}}\\
\vspace{.3in}

Centro Brasileiro de Pesquisas F\'\i sicas - CBPF\\ Rua Dr. Xavier
 Sigaud 150, Rio de Janeiro, RJ, 22290-180, Brazil\\
\vspace{.4cm}

\end{center}
\vspace{0.5in}

\subsection*{Abstract}
The zero-point energy of a massless fermion field in the interior of two 
parallel plates in a $D$-dimensional space-time at zero temperature is 
calculated. In order to regularize the model, a mix between dimensional and 
zeta-function regularization procedure is used and it is found that the 
regularized zero-point energy density is finite for any number of space-time dimensions. We present a general expression for the Casimir energy 
for the fermionic field in such a situation.

\hspace{-.6cm}PACS categories: 11.10.Gh, 11.10.Kk, 12.39.Ba.

\end{titlepage}
\newpage
\baselineskip .37in
Quantum Chromodynamics (QCD) is a non-abelian gauge field theory that
describes the quark dynamics. The non-observation of isolated quarks,
{\it i.e.}, the fact that only colorless states are physically realized,
introduced the concept of confinement. The mechanism by which this occurs
is still unknown. The bag model is an attempt to conciliate the success
of the quark model with confinement and its key point is the approximation 
that the bag is a classical spherical cavity with quarks and gluons confined. 
In a first approximation, the quarks  and gluons move freely inside the bag, 
but are absolutely confined to its interior. Since any quantum field submitted to classical boundary conditions gives rise to the Casimir effect \cite{casimir}
, we expect that confinement will give rise to the Casimir energy of the gluon and quark fields \cite{milton}. Over the past twenty five years many authors calculated the renormalized zero-point energy of different quantum fields in the bag model.
The renormalized vacuum energy due to the scalar field in the bag model was presented by Bender and Hays \cite{bender} and more recently by Romeo \cite{romeo1} and also Bordag {\it et al.} \cite{bordag}. 
The calculation for the case of the gluon field was presented by Milton \cite{milton} and also Romeo \cite{romeo2} and for the fermion field
Milton \cite{milton} and also Bender and Hays \cite{bender} obtained 
the Casimir energy in the bag model.

In the renormalization problem, where we have to extract physically finite 
answers, two different physical situations must be studied. The first one is 
the case where the surface whereon the fields satisfy some boundary conditions divides the manifold into two parts, and the fields are present in both
regions. The second one - the bag model case - is the case where 
the fields are confined to the interior of some region, {\it i.e.}, there 
is no field outside \cite{chodos}. In this second case, since there are no exterior modes, if we use a cut-off or the Green's function method to 
regularize the zero-point energy, divergencies appear and we will be able to absorb them by means of a renormalization procedure only if we introduce {\it ad hoc} contact terms. There are many attempts to solve this problem, but to our knowledge, this question is still open in the literature. Using the Weyl theorem \cite{balian} it it not difficult to show that the asymptotic eigenfrequency distribution of any quantum field confined in a finite volume will present divergent terms proportional to geometric parameters of the region to which the field is confined. To keep in mind the problem, let us consider it for the 
case of the a scalar field.  (Units in which $\hbar=c=1$ will be used
throughout this letter. The metric tensor is taken to be
$\eta_{\mu\nu}=diag(1,-1,...,-1).)$

Let us use the mode sum energy $\left<E\right>_{ren}^{mode}$ as the total
renormalized energy defined by:
\begin{equation}
 \left<E\right>_{ren}^{mode}=\int_0^{\infty}d\omega 
 \frac1{2}\omega \left[N(\omega)-N_0(\omega)\right],
\label{modesum}
\end{equation}
where $\frac1{2}\omega$ is the zero-point energy of each mode, 
$N(\omega)d\omega$ is the number of modes with frequencies between $\omega$ 
and $\omega+d\omega$ in the presence of boundaries and $N_0(\omega)d\omega$
is the corresponding quantity evaluated in empty space. 
If $u_{\omega}(\vec{x})$ are the eigenfunctions of $\nabla^2$ with
eigenvalues $-\omega^2$, it is well known that the asymptotic distribution of eigenfunctions for large $\omega$ is given by
\begin{equation}
 N(\omega)=\frac{V\omega^2}{2\pi}\pm\frac{S\omega}{8\pi}+
 \frac1{2\pi^2}\int_{\partial M}\chi dS +0(\omega^{-2})
\label{asymp}
\end{equation}
for Neumann and Dirichlet boundary conditions if the sign is positive or
negative, respectively. $V$ is the three-volume of $M$, $S$ is the
surface area of $\partial M$ and $\chi$ is the trace of the extrinsic
curvature. Since $N_0(\omega)=\frac{V\omega^2}{2\pi}$, for scalar fields
$\left<E\right>_{ren}^{mode}$ diverges like 
$\frac{\pm S}{16\pi}\int_0^{\infty}\omega^2 d\omega$. For electromagnetic
and spinor fields a similar conclusion can be obtained. 

Baacke and Igarashi \cite{baacke} investigate the structure of the divergencies of the regularized zero-point energy of massive fermions confined in a
spherical cavity, assuming the condition of zero current across the surface.
Considering a massless fermionic field inside and outside a
spherical shell, Milton \cite{milton} showed that there was a cancellation of these divergencies. It is clear that if we consider fermions only in the
interior of the shell, it is necessary to introduce contact terms to deal
with the divergencies of the regularized zero-point energy. 
Thus a question arises: using an analytical regularization procedure,
is it possible to present a geometric configuration 
with confined fields where the contact terms are not necessary?

Our purpose is to compute the zero-point energy due to a free massless
fermionic field confined in the interior of two parallel
plates in any number of space-time dimensions. The same idea was used by Chodos 
and Thorn \cite{chodos}, where the Dirac field is confined between two 
parallel plates separated by a distance $L$ in the $z$ direction - a slab. We 
treat the number of dimensions as a continuous parameter and to regularize the 
zero-point energy we will make use of dimensional regularization in the
continuous variables (related to the sides of the box with infinite length)
and then we analytically extend the Hurwitz zeta-function that appears after
dimensional regularization. As a consequence, we have been able to prove 
that the regularized zero-point energy density is finite in any number of space-time dimensions. In other words, it is not necessary to introduce counterterms in the bare lagrangian, {\it i.e.}, we have a situation with regularization without renormalization. 
Some comments are in order. The bag boundary conditions for spin-half fields 
in rectangular cavities, that is, the 
assumption that the field exists only inside some region, can be implemented
only if there is solely one direction of finite extension, because the 
presence of the corners prevents solutions to the massless Dirac equation
for higher-dimensional hypercubes \cite{ambjorn}. The situation for massive 
spinor fields is even worse, in that there is not a solution even in the 
presence of only one direction of finite size \cite{allcock}.

Consider an ultrastatic $D$-dimensional flat manifold. 
The zero-point energy of the Dirac field is:
\begin{equation}
 \left<0|\hat{H}|0\right>=-c(D)\int\frac{d^d p}{(2\pi)^d}\omega_p,
\label{vac2}
\end{equation}
where $c(D)$ is the number of different spin states in a $D$-dimensional space-time. $c(D)=2^{\frac{D-2}{2}}$, for $D$ even, and 
$c(D)=2^{\frac{D-3}{2}}$ for odd $D$. In the above,
\begin{equation}
 \omega_p^2=p_1^2+p_2^2+...+p_{d-1}^2+p_d^2,
\label{13}
\end{equation}
are the eigenfrequencies of the orthonormal modes, basis in the space of
solutions of the free Dirac equation, {\it i.e.}, these modes are positive
frequency with respect to a timelike Killing vector $\partial_t$. 

We now turn to the calculation of the zero-point energy of the Dirac field
in the $D$-dimensional slab configuration, that is, in the presence of two
parallel plates placed in the $d$-direction, at positions $x_d=0$ and 
$x_d=L$. The boundary condition that we will impose is that there is no 
particle current through the walls, and consequently we will call this
configuration the slab-bag. In Lorentz covariant form we have:
\begin{equation}
 \eta^{\mu}{\overline \Psi}\gamma_{\mu}\Psi=0,  
\label{1}
\end{equation}
where $\eta^{\mu}=(0,\vec{\eta})$, $\vec{\eta}$ being the unit spatial 
vector normal to the surface and directed to the
interior of the slab. The boundary conditions are satisfied only if the
allowed values for the momentum in the $d$-direction are given by 
\cite{milonni}
\begin{equation}
 p_d=p_d(n,L)=(n+\frac1{2})\frac{\pi}{L},\,\,\,\,\,n=0,1,2,3,...
\label{12}
\end{equation}

The zero-point energy of the Dirac field in the slab-bag configuration,
taking into account Eqs. (\ref{13}) and (\ref{12}), and also noting that the integrations over the momenta in Eq. (\ref{vac2}) for the present configuration correspond to $d-1$ integrations and one summation, is given by:
\begin{equation}
\left<0|\hat{H}|0\right>=-c(D)\prod_{i=1}^{d-1}\left(\frac{L_i}{2\pi}\right)
\int\limits_{0}^{\infty}dp_{1}\int\limits_{0}^{\infty
}dp_{2}\,...\int\limits_{0}^{\infty }dp_{d-1}\sum_{n=0}^{\infty }\left(
p_{1}^{2}+...+p_{d}^2(n,L)\right) ^{\frac{1}{2}}.
\label{15}
\end{equation}
Defining the total zero-point energy per unit area of the plates,
that is, the vacuum energy density: 
\begin{equation}
\varepsilon_{D}=\frac{\left<0|\hat{H}|0\right>}
 {\prod L_{i}}=\varepsilon_{d+1}(L),
\label{16}
\end{equation}
we arrive at:
\begin{equation}
 \varepsilon_{D}(L)=-\frac{c(D)}{(2\pi)^{D-2}}\sum_{n=0}^{\infty }
 \int_{0}^{\infty }d^{D-2}p\left(
 p_{1}^{2}+...+p_{d}^{2}(n,L)\right)^{1/2}.
\label{17}
\end{equation}
The expression above is clearly divergent, both in the integration and
in the summation variables. To regularize it we will use dimensional 
regularization in the continuous variables and then analytically extend
the Hurwitz zeta-function that will appear after dimensional regularization.
Using the well-known result of dimensional regularization, {\it i.e.}: 
\begin{equation}
\int \frac{d^{d}u}{\left( u^{2}+a^{2}\right) ^{s}}=\frac{\pi ^{\frac{d}{2}}}{
\Gamma (s)}\Gamma \left(s-\frac{d}{2}\right)
\frac{1}{\left( a^{2}\right) ^{s-\frac{d}{2}
}},
\label{18}
\end{equation}
it is easy to show that the vacuum energy per unit area is given by:
\begin{equation}
\varepsilon _{D}\left(L\right)
=\frac{c(D)\pi^{(D-1)/2}}{2^{D-1}L^{D-1}}\,\Gamma\left(\frac{1-D}{2}\right)
\sum_{n=0}^{\infty}(n+\frac1{2})^{D-1}.
\label{19}
\end{equation}
Next we can define $f(D)=\frac{c(D)\pi^{(D-1)/2}}{2^{D-1}}$, 
and note that:
\begin{equation}
 \zeta(z,q)=\sum_{n=0}^{\infty}\frac1{(n+q)^z},\,\,\,\,\,q\neq 0,-1,-2,-3,...
\label{20}
\end{equation}
is the Hurwitz zeta-function, which is analytic for $Re(z)>1$, in terms of 
which we can write the vacuum energy density of the massless Dirac field in 
the slab-bag configuration as:
\begin{equation}
 \varepsilon_{D}(L)=\frac{f(D)}{L^{D-1}}\Gamma\left(\frac{1-D}{2}\right)
 \zeta\left(1-D,\frac1{2}\right).
\label{21}
\end{equation}
The two cases, {\it i.e.}, $D$ even and $D$ odd, are treated separately.
We first examine the case of even-$D$. For even-$D$, the gamma-function 
presents no poles, and using the principle of analytic continuation
one can write that:
\begin{equation}
 \zeta(-m,v)=\frac{-B_{m+1}(v)}{m+1},\,\,\,\,m=1,2,3,...
\label{22}
\end{equation}
and also the relation between the Bernoulli polynomials and Bernoulli
numbers: $B_D(\frac1{2})=-(1-2^{1-D})B_D$, to obtain the vacuum energy density:
\begin{equation}
 \varepsilon_{D}(L)=-\frac{f(D)}{L^{D-1}}\,\Gamma\left(\frac{1-D}{2}\right)
 (2^{1-D}-1)\,\frac{B_D}{D}\,\,\,\,\,\,for\,D\,even.
\label{23}
\end{equation}
As a conclusion, Eq. (\ref{23}) shows us that the regularized zero-point energy
per unit area of massless fermions in an even-dimensional slab-bag is free of 
divergencies. In particular, calculating the above expression for $D=4$, one 
finds the value:
\begin{equation}
 \varepsilon_{4}(L)=-\frac{7\pi^2}{2880L^3},
\label{D4}
\end{equation}
in accordance with \cite{johnson}.

In the case where $D$ is odd, one begins with the duplication formula
for the gamma functions:
\begin{equation}
 \Gamma\left(-\frac{s}{2}\right)=\frac{\sqrt{\pi}\,2^{s+1}}
 {\Gamma\left(\frac{1-s}{2}\right)}
 \Gamma(-s),
\label{24}
\end{equation}
then defines 
\begin{equation}
 g(D)=\frac{f(D)\sqrt{\pi}\,2^{D}}{\Gamma\left(\frac{2-D}{2}\right)},
\label{25}
\end{equation}
in terms of which the vacuum energy density now reads:
\begin{equation}
 \varepsilon_{D}(L)=\frac{g(D)}{L^{D-1}}
 \,\Gamma(1-D)\,\zeta\left(1-D,\frac1{2}\right).
\label{26}
\end{equation}
Let us now show that the zeta-function times the gamma-function in the 
equation above is a meromorphic function in the whole complex plane.
First we write the integral representation of the Hurwitz zeta-function:
\begin{equation}
 \zeta(z,q)=\frac1{\Gamma(z)}\int_0^{\infty}dt\,t^{z-1}
 \frac{e^{t(1-q)}}{e^t-1},
\label{27}
\end{equation}
and we separate the range of integration as: $[0,1)+[1,\infty)$,
in order to take advantage of the expansion:
\begin{equation}
 \frac{te^{xt}}{e^t-1}=\sum_{n=0}^{\infty}\frac{B_n(x)}{n!}t^n,\,\,\,\,
 0<|t|<2\pi.
\label{29}
\end{equation}
The integral in the second range, {\it i.e.}, $[1,\infty)$, is a regular
function of $z$, because the integrand diverges for small $t$ 
only, and it will be denoted by
\begin{equation}
 h_1(z,q)=\int_1^{\infty}dt\,t^{z-1}
 \frac{e^{t(1-q)}}{e^t-1}.
\label{30}
\end{equation}
Integrating term by term in the first range, and using
Eq. (\ref{30}), we have that:
\begin{equation}
 \zeta(z,q)=\frac1{\Gamma(z)}\left[h_1(z,q)+\sum_{n=0}^{\infty}
 \frac{(-1)^nB_n(q)}{n!}\frac1{z+n-1}\right],
\label{31}
\end{equation}
where we have used that $B_n(1-q)=(-1)^nB_n(q)$. Gathering everything,
we rewrite Eq. (\ref{26}) as:
\begin{equation}
 \varepsilon_{D}(L)=\frac{g(D)}{L^{D-1}}
 \left[h_1\left(1-D,\frac1{2}\right)+\sum_{n=0}^{\infty}
 \frac{(-1)^n\,(2^{1-n}-1)B_n}{n!}\frac1{n-D}\right],
\label{32}
\end{equation}
where we used again that $B_n(\frac1{2})=-(1-2^{1-n})B_n$. We see from 
Eq. (\ref{32}) that there is a pole in the summation for $n=D$,
with residue:
\begin{equation}
 Res\left[\varepsilon_{D}(L)\right]=\frac{g(D)}{L^{D-1}}
 \frac{(-1)^D(2^{1-D}-1)B_D}{D!}.
\label{33}
\end{equation}
But note that for odd-$D$, $D\neq 1$, $B_{D}=0$, and the residue
of the vacuum energy density in the odd-dimensional case vanishes.
Although $B_1=-\frac1{2}$, Eq. (\ref{33}) also states that the residue vanishes
for $D=1$. Eq. (\ref{32}) is general, and it is valid for the even-dimensional 
case as well. Although there also appears a pole for even-$D$, it is 
cancelled by the gamma function in the denominator of Eq. (\ref{25}), and 
Eq. (\ref{23}) is the result, as can easily be checked.
In this way we have proven our assertion that the regularized zero-point
energy of massless fermions in the slab-bag is free of divergencies.

From Eq. (\ref{21}) one could have taken another route and use that
$\zeta(s,\frac1{2})=(2^s-1)\zeta(s)$ and the reflection formula:
\begin{equation}
 \Gamma\left(\frac{s}{2}\right)\zeta(s)=
 \pi^{s-1/2}\,\Gamma\left(\frac{1-s}{2}\right)\zeta(1-s),
\label{reflection}
\end{equation}
valid for all $s$, to find that:
\begin{equation}
 \varepsilon_D(L)=-\frac{c(D)\,(1-2^{1-D})}{2^{D-1}\pi^{D/2}L^{D-1}}\,
 \Gamma\left(\frac{D}{2}\right)\zeta(D).
\label{epsilon}
\end{equation}
The appearance of possible divergent terms becomes hidden automatically
when one uses the reflection formula for zeta functions. 
The result above is finite for all positive $D$, and is always negative, as 
shown in Fig. 1. (In the plot of the energy we left aside the spin degeneracy 
factor $c(D)$). It tends to $-\infty$ either for $D\rightarrow 0$ and for
$D\rightarrow\infty$, taking on the maximum value of $-4.9\times 10^{-6}$
at $D\approx 26.1$. For $D=4$ result Eq. (\ref{D4}) is of course obtained, and 
this value is $7/2$ times the Casimir energy for a scalar field satisfying 
Dirichlet boundary conditions on two parallel plates. The Casimir energy for
the scalar field with Dirichlet boundary conditions as a function of the
number of dimensions was calculated in Ref. \cite{ambjorn}; it is also always 
negative and the maximum occurs exactly at the same value of $D$.

The pressure that the vacuum exerts on the plates is also negative, which
means that it acts tending to approximate them, and it is given by:
\begin{equation}
 -\frac{\partial}{\partial L}\varepsilon_D(L)=
 -\frac{c(D)\,(D-1)(1-2^{1-D})}{2^{D-1}\pi^{D/2}L^{D}}\,
 \Gamma\left(\frac{D}{2}\right)\zeta(D).
\label{pressure}
\end{equation}

In conclusion, in this letter we discuss open questions concerning the
divergent pieces of the regularized zero-point energy of a fermionic field
confined in a finite volume. We show that the regularized zero-point energy 
of massless fermions in the interior of two parallel plates, the slab-bag,
is finite for any number of space-time dimensions. For such, we made use
of dimensional regularization in the continuous momenta, related with the
$d-1$ transversal directions of infinite length of the bag, and then we analytically extended the Hurwitz zeta-function thereby obtained. For even-dimensional space-times the zero-point energy 
is finite, without need of renormalization; in the odd-$D$ case it is shown that the residue of the polar part vanishes identically and, therefore, the usual subtraction of the polar part of a divergent quantity is also not necessary.
Also using an analytic continuation procedure, Dolan and Nash \cite{dolan}
compute the Casimir energy of massless spin zero fields on spheres $S^N$ and find that, for the case of conformal coupling, the Casimir energy vanishes for even-dimensional spheres, presenting no divergencies, and for odd-dimensional spheres the residue of the pole which appears is shown to vanish giving a 
finite value for the Casimir energy. Self-interacting fermions at finite
temperature and density was studied by A\~na\~nos {\it et al.} in  
\cite{ananos}. These authors investigate, at the $\frac1{N}$ (large $N$)
approximation, the behavior of the effective potential of the Gross-Neveu
model. Still using a mix between zeta and dimensional regularization 
procedures, they found that the regularized effective potential is finite for
odd-dimensional space-times. In the case of the Yukawa model, the same kind of
behavior of the regularized effective potential was found \cite{malb}.

One might be tempted to suppose that these results are related with the
work of Svaiter and Svaiter \cite{nami}. These authors proved that the zeta function method is
equivalent to the cut-off method with the subtraction of the polar terms for
a neutral massless scalar field. The basic idea employed by Casimir
\cite{casimir}, who introduced the cut-off method, is that although the 
zero-point energy is divergent, changes in the configuration lead to a finite 
shift in the total energy. In the exponential cut-off method, in order to
evaluate this shift, the total energy is regularized before the subtraction
from the energy of a reference configuration. This total energy is obtained
by adding the regularized zero-point energy of the field inside and outside
the cavity. This approach seems very natural if we are dealing with a system
in which there is field inside and outside the cavity. However, in the bag
configuration, where the field is supposed to exist only inside some region,
this is not a natural approach and we follow the conclusion of the authors
of Ref. \cite{nami} that the Casimir energy evaluated using the analytic
extension of the zeta function throws away these divergent terms.
A natural extension of this work is to calculate the regularized zero-point
energy for the fermion field for different confining geometries, such as the cylinder and the sphere, still using an analytical regularization procedure. 
Now, whether or not the analytical regularization procedure throws away
the polar part of the regularized vacuum energy deserves further investigation.

\section*{Acknowledgements}
This work was supported by the Conselho Nacional de Desenvolvimento
Cientifico e Tecnologico (CNPq).

\newpage

%--------

\begin{figure}[tb]
 \centerline{\epsfysize=6in\epsffile{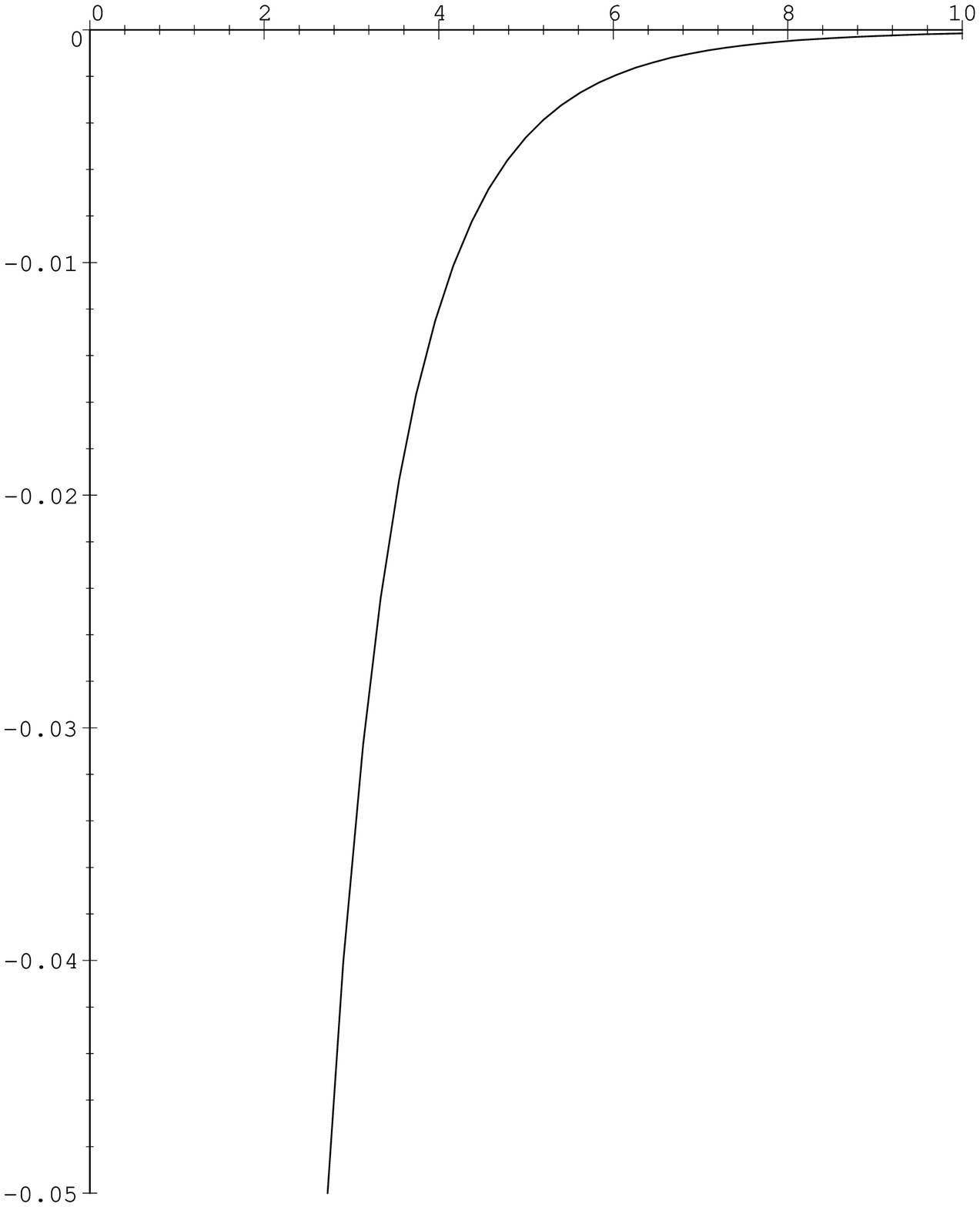}} 
 \caption[region]
 {\small\sf{ The Casimir energy of the spinor field, Eq. (\ref{epsilon}),
 for low $D$, aside the spin multiplicity factor $c(D)$: 
 $e(D)=\frac{\epsilon_D(L)L^{D-1}}{c(D)}$.}}

 \begin{picture}(10,10)

 \put(250,470){$D$}
 \put(85,297){$e(D)$}

 \end{picture}
\end{figure}

\end{document}